\documentclass{easychair}

\usepackage{doc}

\usepackage{graphicx}
\usepackage{url}
\usepackage{amssymb}
\usepackage{amsmath}
\usepackage{paralist}
\usepackage{epstopdf}
\usepackage{balance}
\usepackage{multirow}
\usepackage{hhline}
\usepackage{dcolumn} 
\usepackage[utf8]{inputenc} 
\usepackage{amsmath}

\usepackage{etoolbox}
\apptocmd{\thebibliography}{\footnotesize}{}{}

\newcommand{\uppaal}{\bf{\sc Uppaal}}
\newcommand{\uppaalsmc}{\bf{\sc Uppaal SMC}}

\newcommand{\east}{{\sc East-adl}}

\newcommand{\mats}{{\sc Mats}}

\title{An Energy-aware Mutation Testing Framework for EAST-ADL Architectural Models}

\author{
\small{
Raluca Marinescu\textsuperscript{$\ast$}
\and 
Predrag Filipovikj\textsuperscript{$\ast$}
\and 
Eduard Enoiu\textsuperscript{$\ast$}
\and
Jonatan Larsson\textsuperscript{$\dagger$}
\and 
Cristina Seceleanu\textsuperscript{$\ast$}}
}
\vspace{-8mm}
\institute{
\textsuperscript{$\ast$}\{first.last\}@mdh.se,  \textsuperscript{$\dagger$}jln13010@student.mdh.se\\
M\"alardalen University, V\"aster{\aa}s,  Sweden.
 }

\authorrunning{Marinescu et al. }

\titlerunning{A Resource-aware Mutation Testing Framework}

\begin{document}

\maketitle



\setcounter{tocdepth}{2}
{\small

%
%

\noindent
\textbf{Background and Motivation.}
\label{sec:introduction}
Early design artifacts of embedded systems, such as architectural models, represent convenient abstractions for reasoning about a system’s structure and functionality. 
One such example is the Electronic Architecture and Software Tools-Architecture Description Language ({\east}) \cite{cuenot201011}, a domain-specific architectural language that targets the automotive industry.
{\east} is used to represent both hardware and software elements, as well as related extra-functional information (e.g., timing properties, triggering information, resource consumption). 
Testing architectural models \cite{bertolino2013software} is an important activity in engineering large-scale industrial systems, which sparks a growing research interest. Modern embedded systems, such as autonomous vehicles and robots, have low-energy computing demands, making testing for energy usage increasingly important.
Nevertheless, testing resource-aware properties of architectural models has received less attention than the functional testing of such models.
In our previous work \cite{marinescu2017automatic}, we have outlined a method for testing energy consumption in embedded systems using manually created faults based on statistical model checking of a priced formal system model.
In this paper, we extend our previous work by showing how mutation testing \cite{demillo1978hints} can be used to generate and select test cases based on the concept
of energy-aware mutants--small syntactic modifications in the architectural model, intended to mimic real energy faults. Test cases that can distinguish a certain behavior from its mutations are sensitive to changes in the model, and hence considered to be good at detecting faults. 
The main contributions
of this paper are: (i) an approach for creating energy-related mutants for {\east} architectural models, (ii) a method for overcoming the equivalent mutant problem \cite{madeyski2014overcoming} (i.e., the problem of finding a test case which can distinguish the observable behavior of a mutant from the original one), (iii) a test generation approach based on {\uppaal} Statistical Model Checker {\sc (SMC)} \cite{david2011statistical}, and (iv) a test selection criteria based on mutation analysis using our {\mats} tool\footnote{ {\mats} is an open source software and is available at https://github.com/JLN93/MATS-Tool} \cite{larsson2017automatic}.



  \begin{figure}[tbp]
  	\centering   
    \includegraphics[width=0.72\textwidth]{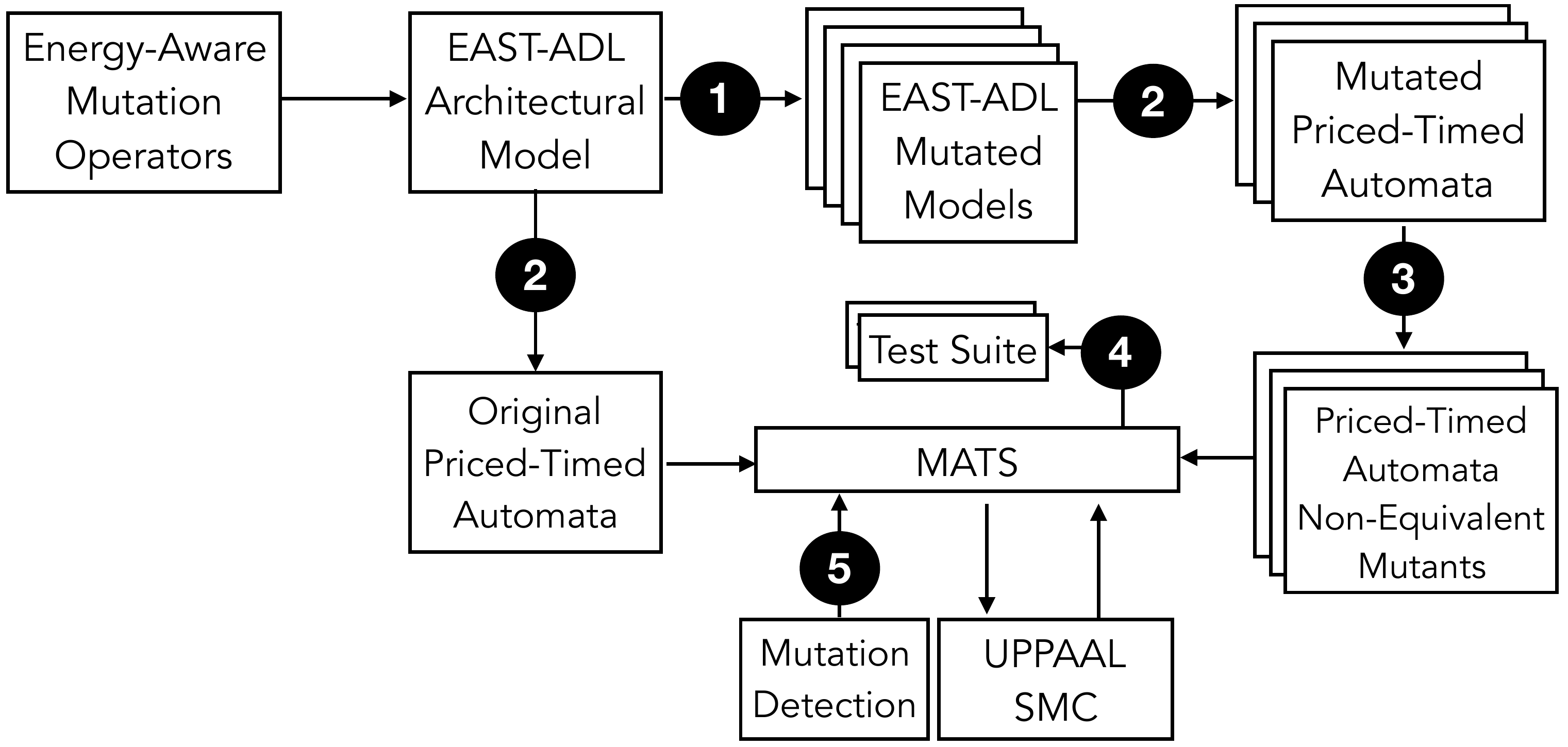}
         \vspace{-0.1cm}
         \vspace{-0.3mm}
 	\caption{Overview of the energy-aware mutation testing framework.}
 	\label{fig:design}
\end{figure}

\vspace{3mm}
\noindent
\textbf{Proposed Framework.}
\label{sec:resource}
In this section, we describe our mutation testing framework that uses energy consumption goals to automatically select test suites based on random system simulations. 
The framework is enabled by transforming the {\east} model into a network of priced timed automata (PTA) \cite{marinescu2015statistical}. It is composed of several steps, mirrored in Figure \ref{fig:design}:

\vspace{2mm}
\noindent
{\sc (1) Energy-Aware Mutant Generation.} 
The consumption of a resource $r$ for an {\east} component represents the accumulated resource usage up to some point in time. 
Based on this assumption, resources can be classified as continuous or discrete \cite{seceleanu2009remes} .
In this paper, we focus on energy consumption, which is a continuous resource assumed to evolve linearly in time ($r(t)=n\times t$, where $n \in \mathbb{N}$ and $t$ is the elapsed time). 
Since in {\east} the resource usage annotation is provided at component level, we define the total energy consumption of the system as $r_{total}(t)=\sum_{i=1}^{m} r_i(t)$, where $m$ is the number of functional components. In mutation testing, faults are injected based on a predefined set of mutation operators. 
Ideally, such mutants should represent commonly-occurring faults, but to the best of our knowledge, there is no previous work on identification of resource-related faults at the architectural level.
Given this, we propose a set of mutation operators applied on: (i) the {\east} resource annotation (Energy Consumption Replacement Operator ({\sc ero})), (ii) the timing behavior of an {\east} component (Period Replacement Operator ({\sc pro}), Execution Time Replacement Operator ({\sc ero})), and (iii) the structure of the functional architecture (Component Removal Operator ({\sc cro}), Component Insertion Operator ({\sc cio}), and Triggering pattern Replacement Operator ({\sc tro})). 
These mutation operators are systematically applied to the entire {\east} model, resulting in a set of energy-aware mutants, each simulating one syntactic model change.

\vspace{2mm}
\noindent
{\sc (2) {\east} to PTA.} In order to use {\uppaalsmc} for test case generation, we transform the {\east} model (with energy consumption annotations) into a PTA model. 
Each {\east} component is automatically transformed into a network of two PTA: an \textit{interface} automaton, which encodes the interface of the component, and a \textit{behavior} automaton, used to model the component's internal behavior.
The triggering of each component, timing information, as well as the resource annotations, are included in the interface PTA. The energy consumption starts at the moment data is read from the input ports until the component writes the data to the output ports. 
This means that the energy consumed by each component increases with the execution time, modeled as a cost ``c'' in PTA ($c(t)=n_c\times t$, where $n_c\in \mathbb{N}$ is the rate of consumption over time $t$), but not when the component is idle ($c^{\prime}(t)=0$). 
A monitor automaton is added to compute the energy used by the system based on the energy consumed by each component. 
For more details, we refer the reader to our previous work \cite{marinescu2015statistical}.

\vspace{2mm}
\noindent
{\sc (3) Detection of Equivalent Mutants.} Let $O_n$ be the original PTA model and $M_m$ be a mutant of the former obtained by applying a predefined mutant operator. We say that models $O_n$ and $M_m$ are \textit{equivalent} if there is no input parameter for which the difference in energy consumption of the models exceeds some predefined threshold within some bounded time limit. Otherwise, there is a valuation of the input parameters for which the mutant can be detected. From the above, it is obvious that the \textit{mutant equivalence check} can be reduced to a \textit{satisfiability problem} \cite{de2011satisfiability}. Let $\Phi = \{\varphi_1, \varphi_2, ..., \varphi_k\}$ and $\Psi = \{\psi_1, \psi_2, ..., \psi_l\}$ denote the set of constraints for the energy consumption of $c_n$ and $c_m$ in $O_n$ and $M_m$, respectively. The mutant $M_m$ is not equivalent to $O_n$ if the following conjunction evaluates to true: $\exists \:l_k. \bigwedge\limits_{i=1}^k \varphi_i(l_k) \land \bigwedge\limits_{j=1}^l \psi_j(l_k) \land (|c_n - c_m| \geq threshold)$, where $l_k$ is an arbitrary input parameter. Reducing the mutant equivalence checking to a satisfiability problem has been considered in other frameworks. Brillout et. al \cite{brillout2010mutation} exploits a similar technique for functional testing of Simulink models. In comparison, our framework is specifically tailored for resource-aware mutation testing of architectural models. The HiLiTe tool \cite{ren2016improving} is another example of using an SMT solver for improving test case generation for large-scale complex and constrained models. Given the fact that the energy consumption is of continuous nature, we have to resort to a specialized type of SMT-solving suitable for hybrid systems \cite{kong2015dreach}.

\vspace{2mm}
\noindent
{\sc (4) Test Suite Generation.} We create executable test cases using the {\mats} tool \cite{larsson2017automatic}, by extracting the input parameters and the energy values at predefined time points from the simulation traces produced by {\uppaal} SMC.
Each test input is a vector of signals where the time-dependent behavior of the model is executed using an ordered sequence of signals. {\mats} uses {\uppaal} SMC for obtaining simulation traces over a predefined number of runs of the system model. A simulation can be formulated as the property:
$simulate\ n [bound] \{ E_1,..,E_k \}$ in {\uppaalsmc},
where $n$ is the number of simulations to be performed, $bound$ is the time bound on the simulations, and $E_1, .., E_k$ are the monitored expressions.
Each test case is executed on both the original model and its mutated counterpart. In order to minimize the final set of test cases we remove the test cases not contributing to the mutation score \cite{larsson2017automatic}.

\vspace{2mm}
\noindent
{\sc (5) Mutant Detection Criteria.} 
We show how to detect energy mutants using the {\mats} tool. 
A mutant is detected by a test suite if the energy signal diverges drastically at certain time points from the expected values (e.g, substantial energy deviations). To measure the mutant-revealing ability of a test suite, we use a quantitative measure of a mutant detection oracle. Let a test case T be generated for a mutated model M, and let
$E_M = {E_{M1}, . . . , E_{MN}}$ be the set of energy signals obtained by running
M for the test inputs in T and sampled at $N$ time points. Let $E_O = {E_{O1},...,E_{ON}}$ be the corresponding expected energy signals. 
We use a threshold to check if the distance between each value of $E_O$ and $E_M$ at each time point is larger than this threshold. If there is at least one energy value in $E_M$ for which the distance is larger than the expected threshold then we consider the mutant M detected. 


\vspace{2mm}
\noindent
\textbf{Conclusions and Future Work.}
\label{sec:con}
In this paper we have outlined a framework for energy-aware mutation testing of {\east} architectural models. Given the large number of energy mutations we aim to reduce the number of equivalent mutants by employing an SMT-solver. In addition, this framework selects test suites contributing to the overall mutation score using {\uppaal} SMC and {\mats}. Future work aims to apply this framework on an industrial case to expose its strengths as well as limitations both in terms of test efficiency and effectiveness.

\vspace{2mm}
\noindent
\textbf{Acknowledgements.} The authors of this work are supported by the following projects:  Swedish Governmental Agency for Innovation Systems (VINNOVA) and ECSEL (EU’s Horizon 2020) under grant agreement No 737494, VINNOVA VeriSpec project 2013-01299, Swedish Research Council (VR) project ``Adequacy-based testing of extra functional properties of embedded systems" and the Swedish Knowledge Foundation (KKS) project DPAC – ``Dependable Platforms for Autonomous systems and Control".
\label{sec:con}

\vspace{-2.5mm}
\label{sect:bib}
\bibliographystyle{plain}
\vspace{-1.5mm}
{\scriptsize
\bibliography{easychair}
}
\end{document}